# A new Majorana platform in an Fe-As bilayer superconductor

Wenyao Liu[1,2,†], Lu Cao[1,2,†], Shiyu Zhu[1,2], Lingyuan Kong[1,2], Guangwei Wang[3], Michał Papaj[4], Peng Zhang[5], Yabin Liu[6], Hui Chen[1,2], Geng Li[1,2], Fazhi Yang[1,2], Takeshi Kondo[5], Shixuan Du[1,7], Guanghan Cao[6], Shik Shin[5], Liang Fu[4], Zhiping Yin[3], Hong-Jun Gao[1,2,7*] and Hong Ding[1,7,8*]

[1]Beijing National Laboratory for Condensed Matter Physics and Institute of Physics, Chinese Academy of Sciences, Beijing 100190, China

[2]School of Physical Sciences, University of Chinese Academy of Sciences, Beijing 100190, China

[3]Department of Physics and Center for Advanced Quantum Studies, Beijing Normal University, Beijing 100875, China

[4]Department of Physics, Massachusetts Institute of Technology, Cambridge, Massachusetts 02139, USA

[5]Institute for Solid State Physics, University of Tokyo, Kashiwa, Chiba 277-8581, Japan

[6]Department of Physics, Zhejiang University, Hangzhou 310027, P. R. China

[7]CAS Center for Excellence in Topological Quantum Computation, University of Chinese Academy of Sciences, Beijing 100190, China

[8]Songshan Lake Materials Laboratory, Dongguan, Guangdong 523808, China

†These authors contributed equally to this work

*Correspondence to: dingh@iphy.ac.cn, hjgao@iphy.ac.cn

**Recently, iron-chalcogenide superconductors have emerged as a new and promising platform[1-3] for studying and manipulating Majorana zero mode (MZM)[4-8]. By combining topological band structure and superconductivity in a multiband material, they provide significant advantages such as higher superconducting transition temperature ($T_c$)[1] and isolated Majorana mode[2,3,9,10]. However, iron-chalcogenide superconductors, especially Fe(Te,Se), suffer from strong inhomogeneity which may hamper their practical application[11]. On the other hand, some iron-pnictide (Fe-As) superconductors, such as LiFeAs, have been demonstrated to have a similar topological band structure[12], yet no MZM has been observed in its vortex cores[13], raising a question of universality of MZM presence in iron-based superconductors. In this work, by using high-resolution angle-resolved photoemission spectroscopy and scanning tunneling microscopy/spectroscopy, we identify the first Fe-As superconductor CaKFe$_4$As$_4$ ($T_c$ = 35 K)[14,15] which has both the superconducting Dirac surface states and the**

**MZMs inside its vortex cores. The topological band inversion is largely due to the down-shift of the $p_z$ band caused by the bilayer band folding [16] in this material. More strikingly, the energies and spatial line profiles of MZM and multiple quantized Caroli-de Gennes-Matricon bound states[17] observed inside the topological vortex can be accurately reproduced by a simple theoretical model derived from a surface Dirac cone, firmly establishing Majorana nature of the zero mode[9].**

The crystal structure of $CaKFe_4As_4$ can be viewed as two different 122-type Fe-As superconductors $CaFe_2As_2$ and $KFe_2As_2$ inserted into one another (Fig.1a)[14]. This special structure not only induces high temperature superconductivity ($T_c$ = 35 K) by self-doping[15], but also breaks the glide-mirror symmetry along the c-axis (Fig. 1a inset). Our density functional theory (DFT) plus dynamical mean field theory (DMFT) calculations (Figs. 1b-d and Fig. 2c) further confirm that the glide-mirror symmetry breaking together with electron correlations create a topological band inversion. The glide-mirror symmetry as seen in $CaFe_2As_2$, is broken due to the difference between Ca and K atoms on the opposite sides of Fe-As layer in $CaKFe_4As_4$[16] (Fig. 1a). Consequently, the Brillouin zone folds along the Γ-Z direction, opening a large hybridization gap (~ 0.5 eV) at the crossing points of the folded $3d_z^2/4p_z$ band, and pushes down its lower branch to below the $d_{xz/yz}$ band at Γ, causing a topological band inversion. Finally, the spin-orbit coupling (SOC) opens a topological gap at the band crossing point between $3d_z^2/4p_z$ and $3d_{xz/yz}$ bands. We note that our DMFT calculation incorporates a mass renormalization of ~ 5 compared to the simple DFT calculation[16], indicating relatively strong correlations, which can reduce the Fermi energy ($E_F$) and coherence length ($\xi_0$) in this material (all the details of our calculations are described in Appendix and Supplementary Materials).

Experimentally, we first performed synchrotron-based angle-resolved photoemission spectroscopy (ARPES) measurements ($h\nu$ = 20 ~ 50 eV, $T_{exp}$ = 18 K). The measured Fermi surfaces (FSs) at $k_z$ ~ π were compared with the ones from our calculation (Fig. 1e). Self-hole-doping effect is reflected by the large areas of hole-like FSs[18]. Band dispersion near high-symmetry points was also measured (Figs. 1f-h), with a comparison to the DFT+DMFT results (Fig.1j), showing a fairly good agreement between the two. Importantly, the innermost hole-like band around Γ has a strong $k_z$ dispersion (Figs. 1f and g), so that the hole-like FS pocket around Z sinks well below the Fermi level at Γ, which is the consequence of the band inversion between $3d_z^2/4p_z$ and $3d_{xz}$ bands. In fact, a clear band dispersion along Γ-Z was observed by our ARPES (Fig. 1i), fully consistent with our band calculations (Fig. 1d).

We next probed the surface state near Γ induced by this topological band inversion using high-resolution laser ARPES ($h\nu$ ~ 7 eV). There is a striking contrast between the broad spectra in the normal state and the sharp spectra in the superconducting (SC) state (Figs. 2a and b, Figs. 2d and e), which is similar to the case of Fe(Te,Se), but quite different from the one in LiFeAs, suggesting that both $CaKFe_4As_4$ and Fe(Te,Se) are bad metals in the normal state[19]. From the two Fermi crossings (#1, #2 in Fig. 2b) near Γ, we can extract two SC gap values (5.9 meV for #1, and 7.5 meV for #2). Since the gap values in most iron-based superconductors, including this material, roughly follow the $\Delta_0\cos(k_x)\cos(k_y)$ formula[20] when measured by ARPES, one might expect that the SC gap at #1 (very close to Γ) would be larger than the one at #2. The fact that the band in the vicinity of Γ has a smaller gap, indicates that this band may come from the surface state within the SOC gap of bulk bands, since the smaller SC gap on the topological surface state is likely induced by the bulk SC gap, just like in the case in Fe(Te,Se)[1-2]. Our surface state calculations also support this scenario, showing a Dirac-cone surface state just slightly above $E_F$ (Fig. 2c).

Encouraged by the promising results from our ARPES measurements and band calculations, we conducted high-resolution (~ 0.3 meV) scanning tunneling microscopy/spectroscopy (STM/S) experiments at low temperature ($T_{exp}$ = 0.45 K) to directly search for the signal of MZM inside a vortex core of this superconductor. A typical cleaved surface shows a good atom-resolved topography revealing that the surface is formed by the As lattice[21], with either Ca or K atoms or clusters scattered on top of it (Figs. 3a and b). We chose a flat region with few cluster of Ca or K on which the STS spectra are homogenous (Fig. 3c). The SC spectra have a main superconducting gap of 5.8 meV, which is likely from the topological surface state with a gap value of 5.9 meV obtained by ARPES, since STM is mostly sensitive to the surface state near Γ. The small bumps at ±3.4 meV are likely coming from the SC gap on the largest hole-like FS since a similar gap value was also observed by ARPES on that FS. The features of SC spectra are similar to the previous results[22,23].

By applying a 2 T magnetic field along the c-axis of the sample, we clearly observed vortex cores on the surface (Fig. 3d). We focused on one vortex within a 20 nm×20 nm area and measured dI/dV spectra along the white arrow across the vortex core (Fig. 3e). A robust zero bias conductance peak (ZBCP) that does not split or shift as it crosses the vortex can be clearly seen in a line-cut intensity plot (Fig. 3f), and a waterfall-like dI/dV spectra plot (Fig. 3g). By utilizing an analytical Majorana wave function derived from Fu-Kane model[2,7], we reproduced the experimental data well with the parameters of the Dirac surface states ($\Delta$ = 5.8 meV, $E_F$ = 20.9 meV, $\xi_0$ = 6.4 nm) estimated from ARPES and calculations.

Besides the zero mode, there are multiple discrete peaks at finite energies inside the vortex core (Fig. 3h). These modes are the quantized Caroli-de Gennes-Matricon bound states (CBSs) under the quantum limit ($T_{ql} \sim T_c\Delta/E_F \sim$ 9.7 K)[9,24,25]. We note that the intensity of CBSs is stronger at the negative energy, consistent with the scenario that the Dirac point is above $E_F$ [13]. As proved recently[9], the CBSs with integer quantized energy levels, coexisting with a robust MZM, are a hallmark of a Dirac-state-induced vortex bound states, since the π Berry phase of a surface Dirac fermion adds additional half-integer level shift[7,26]. To check this behavior, we used a multi-Gaussian fitting to extract the accurate energy positions of discrete bound states inside the vortex (See supplementary Information for more details). We resolved seven discrete levels marked by $L_0$, $L_{\pm 1}$, $L_{\pm 2}$, $L_{\pm 3}$ with the energy of 0 meV, 1.2 meV, 2.5 meV and 3.6 meV, respectively (Fig. 4a). We displayed the extracted energies of the bound states at each spatial position onto the line-cut intensity plot (Fig. 4b), and also did a statistics analysis in a histogram plot (Fig. 4c), showing that the discrete bound states obey an integer quantization at the approximate form of 0 : 1 : 2 : 3. Remarkably, we can use a model calculation using the same parameter set ($\Delta$ = 5.8 meV, $E_F$ = 20.9 meV, $\xi_0$ = 6.4 nm) to reproduce the level energies very well (Fig. 4d).

Due to the high-quality data inside the vortex from this stoichiometric material, we are able to obtain clear spatial distributions for the first four bound states ($L_0$, $L_{-1}$, $L_{-2}$, $L_{-3}$ in Fig. 4e). It shows that the MZM ($L_0$) and the first-level of CBS ($L_{-1}$) have a solid circle with the maximum intensity at the vortex center, while the higher energy CBSs ($L_{-2}$ & $L_{-3}$) show a hollow-ring-like pattern around the vortex center. A numerical calculation using Fu-Kane model[7,9] reproduced our experimental results with good agreement, as the numerical spatial patterns of $L_0$, $L_{-1}$, $L_{-2}$, $L_{-3}$ (Fig.4f) are nearly the same as experimental ones. The direct comparison between the calculated and observed patterns (shown in Fig. 4g) further confirms this. Once again, this calculation used the same set of parameters of the Dirac surface state ($\Delta$ = 5.8 meV, $E_F$ = 20.9 meV, $\xi_0$ = 6.4 nm). Therefore, all the main features of a topological vortex core, including the spatial line profile of MZM, level energies and spatial patterns of discrete CBSs, can be fully reproduced by simple model calculations based on a superconducting Dirac surface state. This establishes in a convincing manner that this high-$T_c$ Fe-As bilayer superconductor, which induces glide-mirror symmetry breaking[16], can host isolated Majorana zero modes on its surface, offering a new and more practical platform for exploring the properties of MZMs and manipulating them.

**Materials and methods**

Single crystals of $CaKFe_4As_4$ were grown using the self-flux method, and the value of $T_c$ was determined to be 35 K from magnetization measurements. Clean surfaces for the ARPES measurements were obtained by cleaving the samples in situ in an ultrahigh vacuum better than 5 × $10^{-11}$ Torr. Synchrotron-based ARPES measurements were performed at the 'Dreamline' beamline of the Shanghai Synchrotron Radiation Facility (SSRF) with a Scienta Omicron DA30L analyzer with energy resolution of ~10 meV. High-resolution laser-ARPES measurements were performed at the Institute for Solid State Physics at the University of Tokyo on an ARPES system with a VG-Scienta HR8000 electron analyzer and a vacuum ultraviolet laser of 6.994 eV, with the energy resolution of ~ 3 meV. The samples used in STM experiments were cleaved *in-situ* and immediately transferred to a STM head. Experiments were performed in ultrahigh vacuum (1 × $10^{-11}$ mbar) low-temperature (T ~ 0.45 K) STM systems of USM-1300-$^3$He with a vector (9 T-2 T-2 T) magnet. The configurations of the system are the same as in the previous work[2,9].

# Acknowledgements

We thank Yaobo Huang, Shunye Gao and Jerry Huang for technical assistance on synchrotron-based ARPES measurements. This work at IOP is supported by grants from the Chinese Academy of Sciences (XDB28000000, XDB07000000), the Ministry of Science and Technology of China (2015CB921000, 2015CB921300, 2016YFA0202300), and the National Natural Science Foundation of China (11888101, 11234014, 11574371, 61390501). The band calculations used high performance computing clusters at BNU in Zhuhai and the National Supercomputer Center in Guangzhou, and Z.P.Y is supported by NSFC (11674030), the Funds for the Central Universities (310421113) and the National Key Research and Development Program of china (2016YFA0302300). L.F. is supported by US DOE (DE-SC0010526). Laser-APRES work was supported by the JSPS KAKENHI (Grants No. JP18H01165, JP19F19030 and JP19H00651). G.H.C is supported by Funds for the Central Universities and the National Key Research and Development Program of china (2019FZA3004, 2017YFA0303002 and 2016YFA0300202).

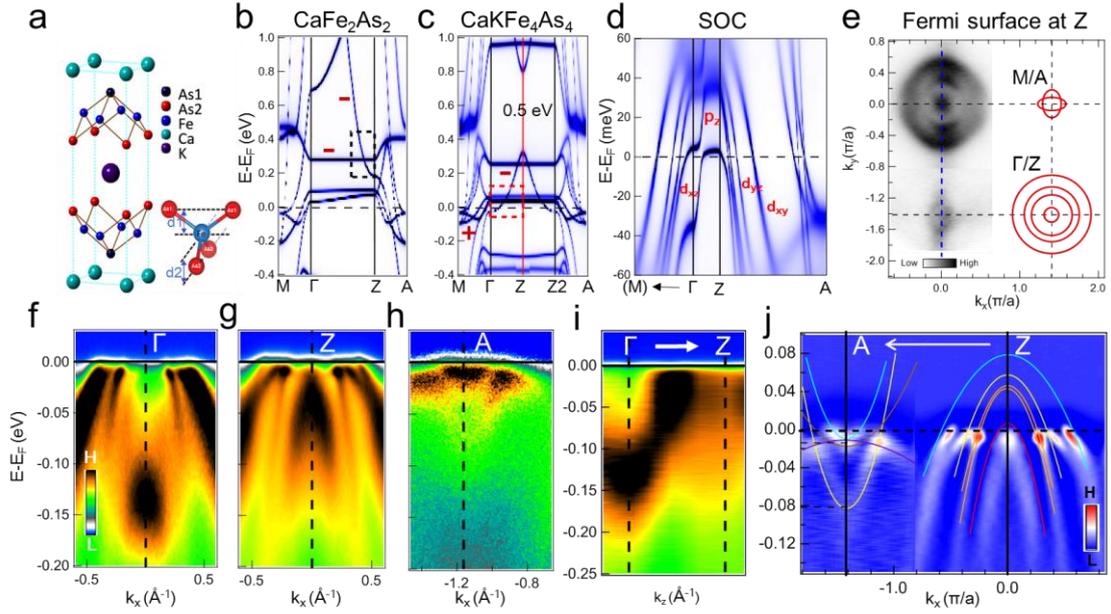

**Figure 1 | Crystal and Electronic Structures for CaKFe$_4$As$_4$.** a, The crystal structure of CaKFe$_4$As$_4$ with the inset of the As-Fe-As tetrahedron. The bond lengths ($d_1$, $d_2$) and angles are different between Fe-As$_1$ and Fe-As$_2$[16].  b-d, DMFT+DFT calculation results for CaFe$_2$As$_2$, CaKFe$_4$As$_4$, and CaKFe$_4$As$_4$ with SOC, respectively. The glide-mirror symmetry breaking effect in 1144 system is visible by comparing the band structures of (b) CaFe$_2$As$_2$ and (c) CaKFe$_4$As$_4$. A large hybridization gap (~ 0.5 eV) is formed between the folded p$_z$ bands in CaKFe$_4$As$_4$. A band inversion with a SOC gap of 20-30 meV is found near $E_F$ at (d). The red symbols of "+" and "-" represent the band parity.   e, Fermi surfaces measured by ARPES at a photon energy of 35 eV (close to the Z point), the red contours on the right represent the predicted bulk Fermi surfaces calculated by DMFT+DFT.   f-h, ARPES spectral intensity plots along the blue dashed line in (e) with a $p$-polarized at photon energies of 42 eV (Γ) and 33 eV (Z and A).   i, ARPES spectral intensity plot along the Γ-Z direction measured under photon energies from 21 – 45 eV.   j, The second derivative of ARPES intensity plot along the A-Z direction obtained from (g) and (h), with comparison to the calculated results from (d) (plotted as colored lines).

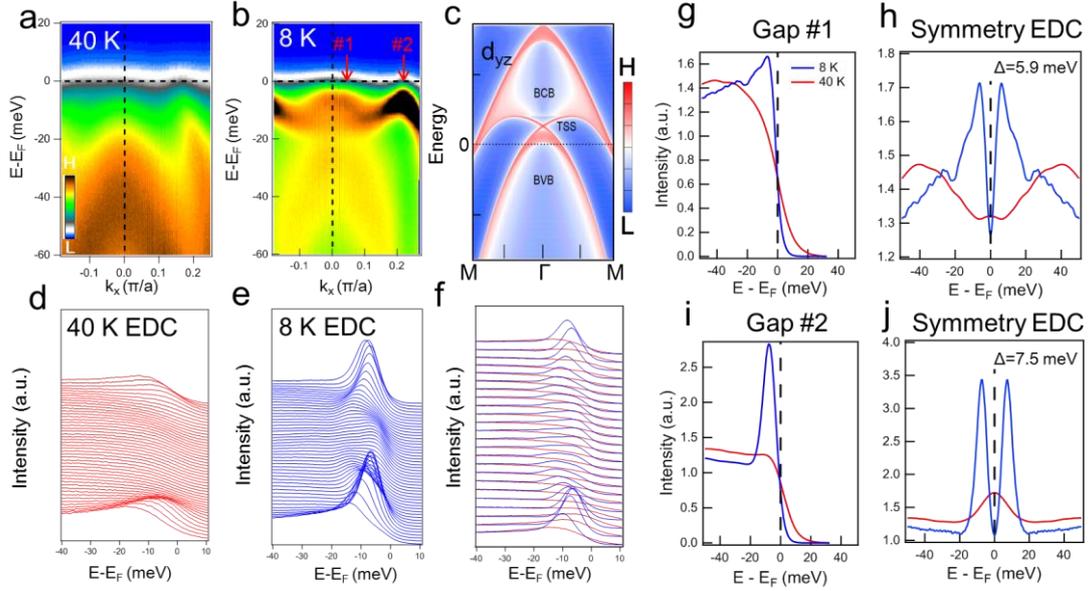

**Figure 2 | Laser-ARPES results and calculated results for the topological surface state.** a, ARPES spectral intensity plot along GM, measured with a *p*-polarized 7-eV laser at 40 K. b, Same as (a) but at 8 K. c, The calculated band structure projected onto the (001) surface. The topological surface state (TSS) along with the bulk valence band (BVB) and bulk conduction band (BCB) are plotted. d and e, Energy distribution curves (EDCs) at 40 K and 8 K, respectively. f, Overlap of EDCs of 40 K (red curves) and 8 K (blue curves) from (d) and (e). g-j, EDCs (g, i) and the symmetrized EDCs (h, j) at $k_F \sim 0.04$ π/a and 0.2 π/a (indicated by #1, #2 in (b)) measured at 8 K (blue) and 40 K (red). Superconducting gap values (5.9 meV in (h), and 7.5 meV in (j)) can be estimated from the two sharp peaks in the symmetrized EDCs. We attribute the gap of 5.9 meV to the SC gap in the topological surface band.

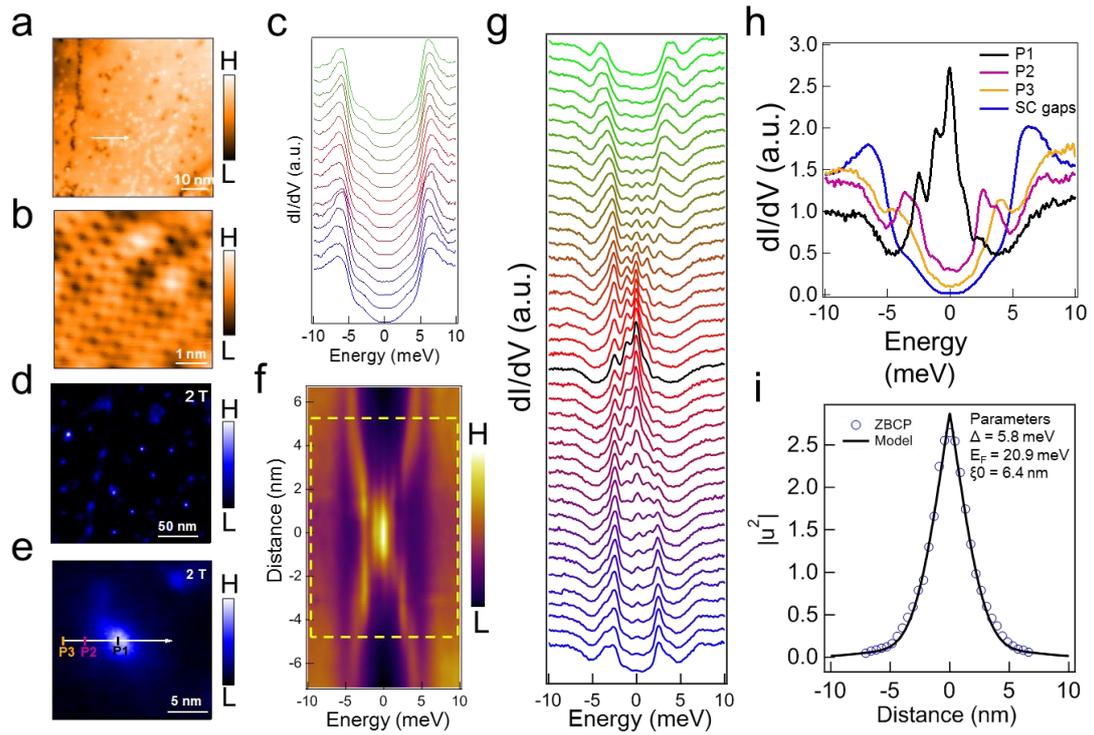

**Figure 3 | STM Topography, vortex mapping, superconductor gap, and energetic and spatial profile of vortex bound states.** a, STM topography of CaKFe$_4$As$_4$ (scanning area: 70 nm × 70 nm). b, An atom-resolved topography within the area of 5 nm × 5 nm. c, STS dI/dV spectra taken along the white arrow in (a), at T = 0.4 K and B = 0 T. d, Normalized zero-bias conductance map measured at a magnetic field of 2 T in the area as shown in (a). e, Zero-bias conductance map (area: 20 nm × 20 nm) around a vortex core at T = 0.45 K and B = 2 T. f, STS intensity plot of dI/dV spectra along the white line across the vortex in (e). g, Waterfall-like dI/dV spectra plot of (f) within the yellow dashed box in (h), with the black curve representing the spectrum at the vortex core center. h, Comparison of dI/dV spectra at vortex core (P1), middle (P2), edge (P3), and without the magnetic field. i, Comparison between the spatial dependence of the ZBCP peak height with a theoretical calculation of the MZM spatial profile (to exclude the influence of CBSs and background, each STS spectrum is fitted by multi-Gaussian peaks to extract the height of ZBCP. An example of a fit is shown in Fig. 4).

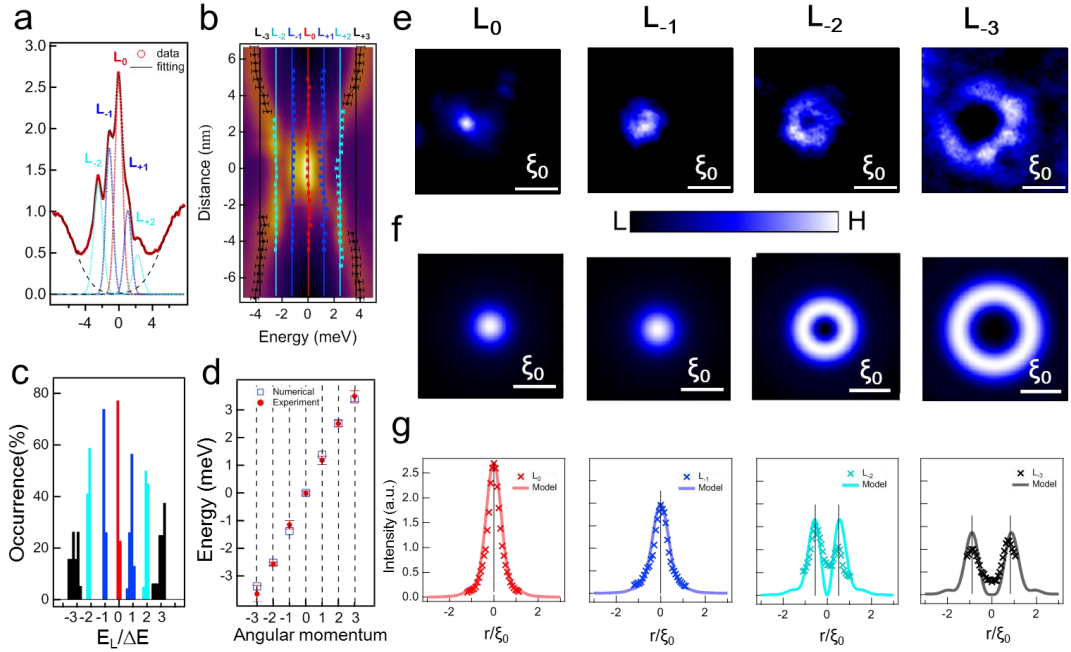

**Figure 4 | Integer quantized energy levels and spatial profiles of MZM and CBSs.**
a, Multiple Gaussian fit for the STS spectrum at the vortex center of Fig. 3f. The red dots are the experimental data, the colored dashed curves are the fitting curves of the in-gap bound states, and the black solid curve is the final fitting result.   b, The line-cut intensity plot same as the Fig. 3d, with the colored marked lines representing the ZBCP and discrete quantized CBSs at different energies marked by $L_0$, $L_{\pm 1}$, $L_{\pm 2}$, $L_{\pm 3}$, which are obtained from the multiple Gaussian peak fits for all the spectra in Fig. 3g. The solid line is calculated using $E_L/\Delta E_{average} = n$ ($\Delta E_{average}$ = 1.23 meV is the average value of energy level spacing), while $n$ is the number of the energy level.   c, A histogram of the energy values of all the observed in-gap states with a sampling width of 50 μeV. The horizontal energy scale is normalized by the first level spacing. d, Comparison between the numerically calculated energy eigenvalues of CBSs and experimental values for different angular momenta.   e, Spatial patterns of vortex bound states at voltage bias equal to 0, -1.2, -2.4, and -3.6 mV, respectively. The size of the area is scaled by the coherence length $\xi_0$.   f, Numerical calculations of the two-dimensional local density of $L_0$ (MZM), $L_{-1}$, $L_{-2}$ and $L_{-3}$, respectively, which are based on the topological vortex core model. In this calculation, the integer-quantized CBS levels are located at -1.38 meV, -2.52 meV, and -3.38 meV, respectively.   g, Comparison between the wave functions of STM results (crosses) and numerical calculations (lines). The experimental spatial profiles of energy states are extracted along the solid lines in (b). The intensity of numerical local density of states is rescaled to be comparable to the experimental data.